\documentclass[conference,10pt]{IEEEtran}
\IEEEoverridecommandlockouts
\usepackage{cite}
\usepackage{amsmath,amssymb,amsfonts}
\usepackage{algorithmic}
\usepackage{graphicx}
\usepackage[font=footnotesize]{caption}
\usepackage{subcaption}
\usepackage{textcomp}
\usepackage{xcolor}
\usepackage{multirow}
\usepackage{tikz}
\usepackage[hidelinks]{hyperref}
\usepackage{stfloats}

\newtheorem{definition}{Definition}

\definecolor{mypurple}{rgb}{0.4, 0.2, 0.6}

\newcommand\encircle[1]{%
  \tikz[baseline=(X.base)] 
    \node (X) [draw, font=\small, align=center, shape=circle, inner sep=-0.2ex, scale=1, fill=black, text=white] {\strut #1};%
}
\newcommand\encirclepurple[1]{%
  \tikz[baseline=(X.base)] 
    \node (X) [draw=mypurple, line width=0.2mm, font=\small, align=center, shape=circle, inner sep=-0.2ex, scale=1, fill=mypurple, text=white] {\strut #1};%
}

\def\BibTeX{{\rm B\kern-.05em{\sc i\kern-.025em b}\kern-.08em
    T\kern-.1667em\lower.7ex\hbox{E}\kern-.125emX}}

\usepackage{cite}

\begin{document}

\title{SARA: A Stall-Aware Memory Allocation Strategy for Mixed-Criticality Systems}


\author{\IEEEauthorblockN{Meng-Chia Lee\IEEEauthorrefmark{1}\IEEEauthorrefmark{2}, Wen Sheng Lim\IEEEauthorrefmark{2}, Yuan-Hao Chang\IEEEauthorrefmark{1}\IEEEauthorrefmark{2}, Tei-Wei Kuo\IEEEauthorrefmark{2}}
\IEEEauthorblockA{\IEEEauthorrefmark{1}Institute of Information Science, Academia Sinica, Taipei, Taiwan \\ 
\IEEEauthorrefmark{2}Department of Computer Science and Information Engineering, National Taiwan University, Taipei, Taiwan \\ 
r11922052@ntu.edu.tw, tundergod1882@gmail.com, johnson@csie.ntu.edu.tw, ktw@csie.ntu.edu.tw}
}


\maketitle

\begin{abstract}
    The memory capacity in edge devices is often limited due to constraints on cost, size, and power. Consequently, memory competition leads to inevitable page swapping in memory-constrained mixed-criticality edge devices, causing slow storage I/O and thus performance degradation. In such scenarios, inefficient memory allocation disrupts the balance between application performance, causing soft real-time  (soft RT) tasks to miss deadlines or preventing non-real-time (non-RT) applications from optimizing throughput. Meanwhile, we observe unpredictable, long system-level stalls (called \textit{long stalls}) under high memory and I/O pressure, which further degrade performance. In this work, we propose a Stall-Aware Real-Time Memory Allocator (SARA), which discovers opportunities for performance balance by allocating just enough memory to soft RT tasks to meet deadlines and, at the same time, optimizing the remaining memory for non-RT applications. To minimize the memory usage of soft RT tasks while meeting real-time requirements, SARA leverages our insight into how latency, caused by memory insufficiency and measured by our proposed PSI-based metric, affects the execution time of each soft RT job, where a job runs per period and a soft RT task consists of multiple periods. Moreover, SARA detects long stalls using our definition and proactively drops affected jobs, minimizing stalls in task execution. Experiments show that SARA achieves an average of 97.13\% deadline hit ratio for soft RT tasks and improves non-RT application throughput by up to 22.32$\times$ over existing approaches, even with memory capacity limited to 60\% of peak demand.

\end{abstract}


\section{Introduction} \label{sec:intro}
    Due to constraints on cost, size, and power, edge devices are often equipped with limited memory capacity~\cite{larimi2021understanding, wu2023app, capotondi2020cmix, chang2021survey, deng2020edge, weiner2022tmo, xue2011emerging}. In such memory-limited edge devices, competition for memory among processes leads to frequent page swapping and subsequent slow storage I/O~\cite{tavakkol2018mqsim, wu2020joint}, resulting in performance degradation. This highlights the importance of efficient memory allocation to meet application performance requirements. In particular, edge AI applications impose diverse requirements, including both soft real-time (soft RT) applications for responsive user interactions and non-real-time (non-RT) applications for substantial local data storage and processing capabilities~\cite{cinque2022virtualizing, cinque2019rt}. This is because the proliferation of edge AI applications has not only raised user expectations for low-latency inference but also emphasized strong privacy guarantees~\cite{sachdev2020towards, mao2023security}, requiring edge devices to provide efficient services while processing and managing large volumes of privacy-sensitive data locally. In such mixed-criticality systems, inappropriate memory allocation fails to \textit{balance the trade-off between real-time guarantees and throughput}, potentially causing soft RT tasks to miss deadlines or preventing non-RT applications from optimizing throughput.


    Memory allocation strategies based on Linux control groups (cgroups) are widely adopted to manage page-swapping overhead~\cite{delimitrou2014quasar,sfakianakis2018quman, zhu2022qos,lagar2019software,weiner2022tmo}. Rather than modifying the Linux page replacement policy to retain specific pages in memory~\cite{lee2012mrt,wu2023app}---which incurs high management overhead and lacks adaptability to OS updates---these strategies preserve modular OS design (without OS-level modification). They achieve this by actively controlling the memory space allocated to each application through cgroups in user space.

    Among memory allocation approaches using cgroups, current enterprise solutions typically employ greedy approaches that allocate excessive memory to soft RT applications~\cite{memory_sufficiency}, ensuring their timing requirements but at the cost of non-RT application throughput. In contrast, offline profiling~\cite{delimitrou2014quasar, sfakianakis2018quman, zhu2022qos} estimates the memory space required to achieve expected process performance, but does not account for runtime variability (e.g., jitter). Lagar-Cavilla et al.~\cite{lagar2019software} use the online metric to maintain a constant page swapping-in rate for runtime memory allocation, but overlook latency variations of different storage devices, potentially resulting in missed deadlines for soft RT tasks on slower storage. TMO~\cite{weiner2022tmo} proposes Pressure Stall Information (PSI) to measure the performance impact of memory scarcity, regardless of the type and speed of storage.
    However, it uses PSI primarily for memory space allocation in data centers to optimize throughput, rather than for mixed-criticality systems.

    Additionally, another challenge arises from our experimental observation that memory-constrained systems occasionally experience significant and unpredictable \textit{system-level stalls} (termed "\emph{long stalls}"), which delay application execution and cause cascading effects, resulting in soft RT task deadline misses and degraded non-RT application throughput.

    In this work, we introduce a Stall-Aware Real-Time Memory Allocator (SARA), a novel real-time memory management strategy designed to allocate memory space for mixed-criticality applications on memory-constrained edge devices. SARA aims to satisfy soft RT task timing requirements while optimizing non-RT application throughput. Our main contributions are as follows. First, we identify opportunities for performance balance by (1) allocating just enough memory to soft RT tasks to meet deadlines while optimizing the remaining memory for non-RT applications, and (2) proposing a novel PSI-based metric, $s_{intv}$, to observe how latency caused by memory shortages affects the execution time of each soft RT job, where a job runs per period and a soft RT task consists of multiple periods (Section~\ref{sec:app:first}). Second, based on this observation, we derive appropriate $s_{intv}$ thresholds to assess memory sufficiency and efficiency for soft RT tasks to meet deadlines and dynamically adjust memory allocation to optimize non-RT application throughput (Section~\ref{sec:app:second}). Third, we formally define long stalls through a systematic analysis of $s_{intv}$ and implement early job dropping for soft RT tasks when long stalls are detected to mitigate their impact on application performance (Section~\ref{sec:app:third}).

    SARA was implemented on Linux kernel version 6.5.0 with cgroup v2 and evaluated using both fast and slow SSDs as swap backends to demonstrate its effectiveness across different hardware configurations. Compared to existing memory management strategies, SARA achieves a high deadline hit ratio of 97.13\% on average for soft RT tasks while simultaneously improving non-RT application throughput by up to 22.32$\times$. Notably, we maintain these benefits even with memory capacity limited to 60\% of peak demand.

\section{Background and Motivation} \label{sec:back_motv}
    \subsection{Impacts of Memory Insufficiency} \label{sec:back:knowledge_1}
    
    \begin{figure}[hbt]
        \centering
        \includegraphics[width=0.33\textwidth]{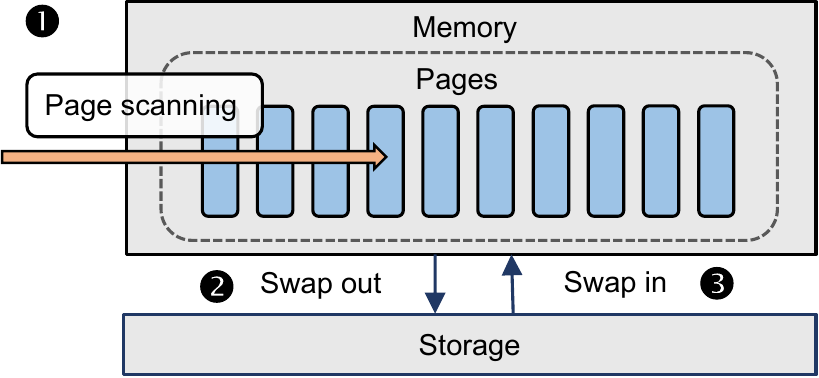}
        \caption{Linux's Page Replacement Policy.}
        \label{fig:page_reclamation_and_IO_overhead}
    \end{figure}

    When memory usage reaches its limit, the Linux kernel employs a sophisticated page replacement mechanism to free up memory space by moving pages that are less frequently accessed to secondary storage (e.g., SSD)~\cite{wu2023app}. As illustrated in \figurename~\ref{fig:page_reclamation_and_IO_overhead}, when a page fault occurs and there is insufficient memory space to accommodate the faulting page, the OS scans pages to identify those that are infrequently accessed for reclamation~\encircle{1}~\cite{Reclamation}. The selected pages are swapped out to storage devices~\encircle{2} and makes room for the faulting page to be swapped into memory~\encircle{3}.

   \begin{figure}[hbt]
        \centering
        \includegraphics[width=0.28\textwidth]{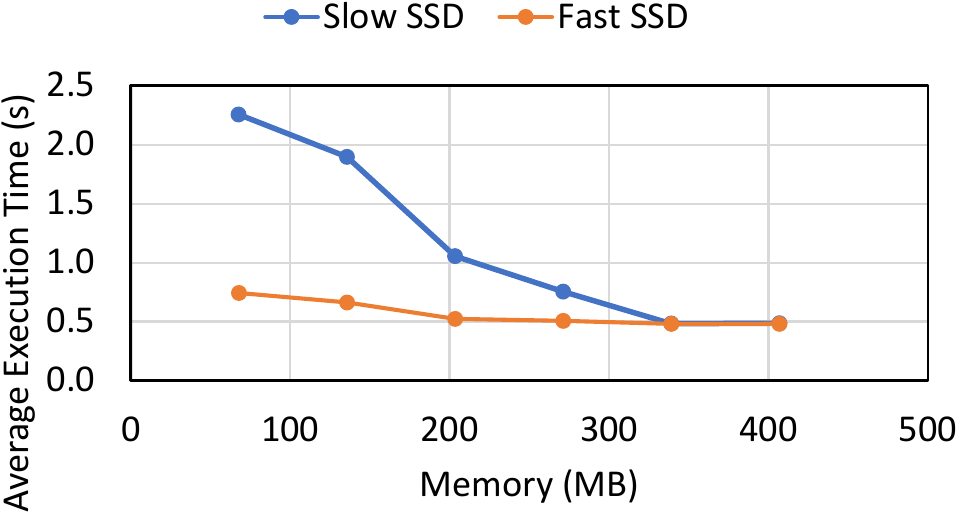}
        \caption{Impact of memory allocation on application performance.}
        \label{fig:performance_vs_memory}

    \end{figure}

    It is important to note that the page replacement process involves storage I/O operations, which typically incur orders of magnitude higher latency compared to memory accesses~\cite{tavakkol2018mqsim, wu2020joint}. This can introduce significant overhead to application performance. \figurename~\ref{fig:performance_vs_memory} shows this performance impact on a soft RT periodic task (Sphinx) under varying memory-constrained configurations on systems with slow and fast SSD. With less memory, the average execution time of soft RT jobs over 500 periods increases, raising the risk of deadline misses, whereas on the fast SSD, the increase is slower due to its higher I/O bandwidth. Conversely, increasing memory improves performance. A similar trend is observed for non-RT applications, where less memory results in lower throughput, with the effect being more pronounced on slow SSDs. Therefore, this highlights the importance of proper memory allocation for meeting performance requirements.

    \subsection{Pressure Stall Information (PSI)}

    As shown in \figurename~\ref{fig:performance_vs_memory}, under the same memory allocation, the execution time of the same process may vary depending on the SSD performance. Specifically, the same level of memory scarcity can cause varying degrees of performance degradation based on the SSD's capabilities. Therefore, instead of using metrics like page swapping-in rate~\cite{lagar2019software}
    , we need a metric that reflects performance degradation in relation to SSD characteristics to determine memory allocation.

    \begin{figure}[hbt]
        \centering
        \includegraphics[width=0.30\textwidth]{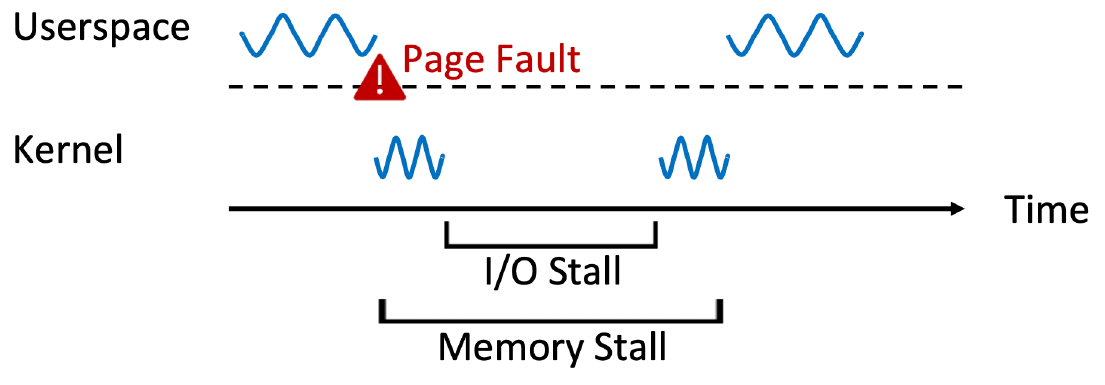}
        \caption{PSI metric.}
        \label{fig:memory_and_io_stall}
    \end{figure}
    
    We identify the potential of Pressure Stall Information (PSI), which was proposed in TMO~\cite{weiner2022tmo}, to capture this varying performance impact of memory scarcity. As illustrated in \figurename~\ref{fig:memory_and_io_stall}, PSI measures the stall time experienced by processes within a \textit{cgroup} after a page fault occurs, including memory stalls and I/O stalls\footnote{CPU stall is negligible in memory-constrained conditions, so we focus on memory and I/O stalls in this paper for clarity.}. Memory stalls include the time spent on memory reclamation and allocation, I/O operations to reload a page, and reading a page from the swap device. I/O stalls measure the time that processes spend waiting for block I/O completion. PSI aggregates the stall time and reports it in two forms: a percentage and an absolute stall time in $\mu s$. The percentage represents the proportion of non-idle processes stalled within an observation window. For example, if the only process in a cgroup experiences 10,000 $\mu s$ of stall time within a 10-second window, this corresponds to a 0.1\% stall. Since PSI directly measures the latency caused by memory shortage, it can capture the varying performance impact across different SSDs. However, no previous work has explored how to use PSI to assess the performance of soft RT tasks. Without understanding this relationship, it can result in inefficient memory allocation, as discussed in Section~\ref{sec:motv:motivation}.

    \subsection{Challenges in Real-Time Memory Management} \label{sec:motv:motivation}

    Memory resource competition between concurrent processes poses significant performance challenges, especially in memory-constrained, mixed-criticality edge systems. As explained in Section~\ref{sec:back:knowledge_1}, memory usage significantly affects process performance. If non-RT applications consume excessive memory, they can severely reduce the memory available to soft RT tasks, thereby increasing the risk of missed deadlines. Conversely, if the execution time of the soft RT job per period is minimized by aggressively allocating memory to it, non-RT application throughput decreases. While existing works attempt to address this issue, none leverage PSI, which can capture the performance impact of memory contention across different SSDs. However, designing memory management approaches that utilize PSI remains challenging for two reasons, which are described below.

    \begin{figure}[hbt]
         \centering
         \begin{subfigure}[t]{0.266\textwidth}
             \centering
             \includegraphics[width=\textwidth]{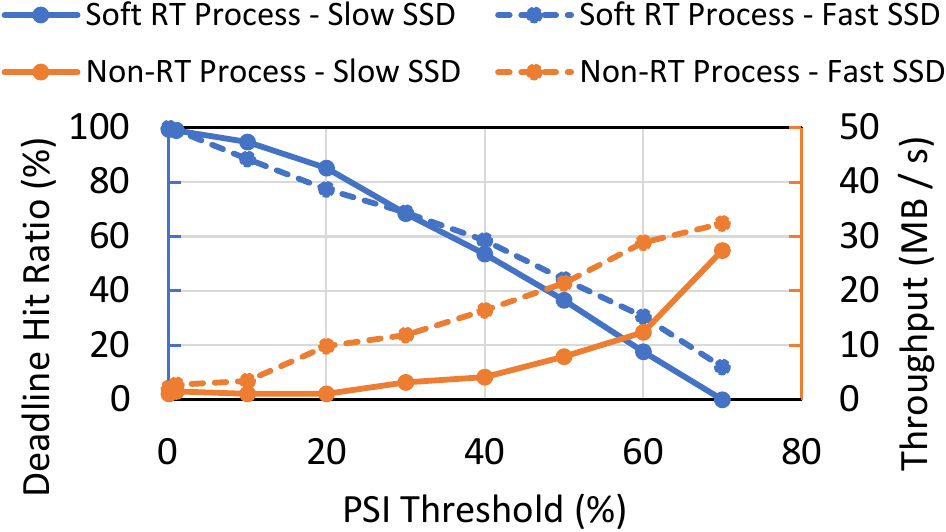}
             \caption{Trade-off between application performance.}
             \label{fig:motv_1.1}
         \end{subfigure}
         \hfill
         \begin{subfigure}[t]{0.216\textwidth}
             \centering
             \includegraphics[width=\textwidth]{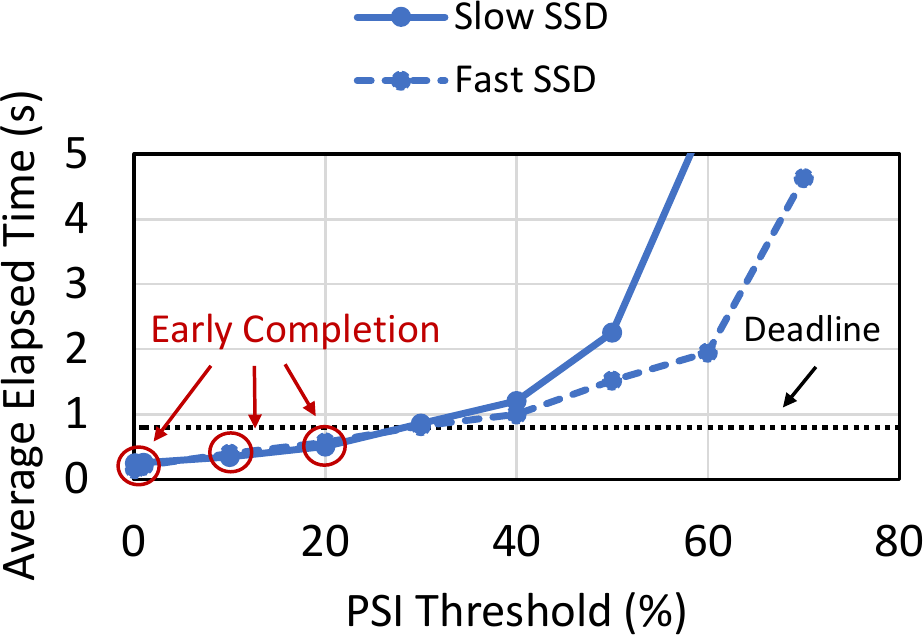}
             \caption{Average job elapsed time under varying PSI thresholds.}
             \label{fig:motv_1.2}
         \end{subfigure}
        \caption{Examples of inefficient memory management.}
        \label{fig:motv_1}
    \end{figure}

    \textbf{Balancing trade-off between real-time and throughput.}
    \figurename~\ref{fig:motv_1} illustrates our investigation of PSI-based memory allocation in systems with fast and slow SSDs. We execute a soft RT periodic task for 500 periods concurrently with a non-RT application. To explore the effects of PSI, we systematically vary the PSI threshold for the soft RT task in 10\% intervals. These thresholds represent the maximum level of performance degradation that the soft RT task can tolerate. We monitor PSI in each time period, and if it falls below the threshold, memory allocation for the soft RT task is reduced, freeing up memory for the non-RT application. \figurename~\ref{fig:motv_1.1} shows the performance trade-off: lower PSI thresholds (below 20\%) lead to increased memory allocation for the soft RT task, improving its deadline hit ratio but reducing non-RT application throughput, and vice versa. This trade-off becomes more pronounced with a fast SSD backend since higher I/O bandwidth allows the PSI of the soft RT task to stay within the thresholds with less memory. As a result, more memory is freed up for the non-RT application, which also benefits from increased I/O bandwidth. However, no single threshold achieves \textit{an optimal balance}. \figurename~\ref{fig:motv_1.2} shows that under low thresholds (below 20\%), soft RT jobs in most periods complete earlier than their deadlines due to memory over-utilization. In fact, these jobs could still meet deadlines with less memory, allowing more memory to be allocated to the non-RT application to optimize throughput. Furthermore, a static threshold results in \textit{unstable elapsed time}. For instance, at a 30\% PSI threshold, while the average elapsed time is close to the deadline, the deadline hit ratio is only 69\%, indicating that some missed-deadline jobs are experiencing memory under-utilization. These two observations imply the need to find \textit{an appropriate threshold} that can accurately manage the performance of soft RT tasks to balance the performance trade-off. Additionally, it should be \textit{dynamically} adjusted at runtime.

    \begin{figure}[hbt]
        \centering
        \includegraphics[width=0.35\textwidth]{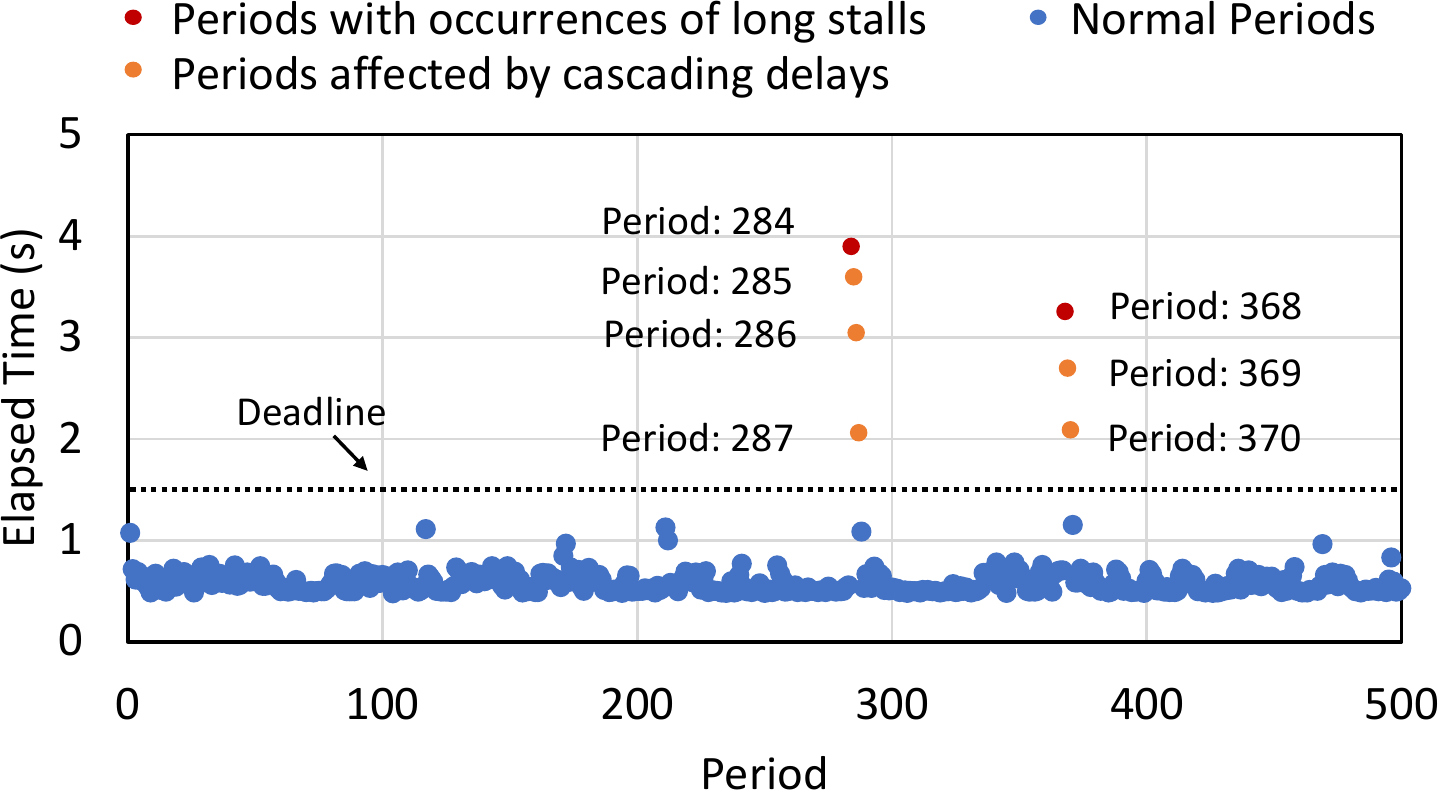}
        \setcounter{figure}{4}
        \caption{The impact of long stalls.}
        \label{fig:motivational_experiment_long_stall}
    \end{figure}

    \textbf{Mitigating long system-level stall.}
    When processes contend for scarce memory resources, we observe \textit{unpredictable and severe} "long system-level stalls" (termed \textit{long stalls}). These stalls may result from transient peaks in overall resource demand. During such stalls, all user-level processes become suspended, causing significant performance degradation.
    \figurename~\ref{fig:motivational_experiment_long_stall} illustrates this phenomenon by showing the elapsed time of jobs over 500 periods of a soft RT periodic task (Sphinx) when co-running with a non-RT application (Graphchi). The long stalls significantly prolong job elapsed times (marked in red), creating cascading delays that affect subsequent jobs (marked in orange) and cause a series of deadline misses. This effect extends beyond soft RT tasks and also degrades the throughput of non-RT applications.
    These observations highlight the critical need for the \textit{timely} detection and mitigation of unpredictable long stalls to optimize application performance.

\section{The proposed SARA} \label{sec:app}   
    \begin{figure}[hbt]
        \centering
        \includegraphics[width=0.48\textwidth]{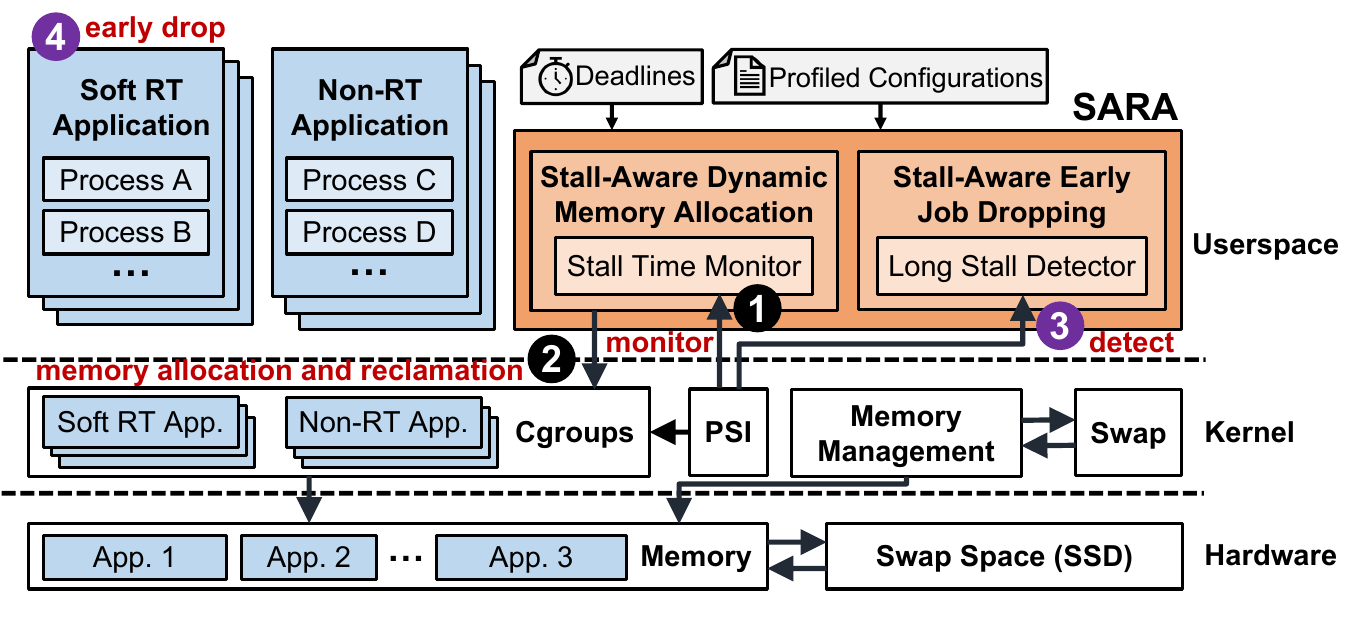}
        \setcounter{figure}{5}
        \caption{Overview of the SARA system architecture.}
        \label{fig:approach_overview}
    \end{figure}  
    \subsection{Overview} \label{sec:app:overview}  
    We propose SARA, a memory management strategy that efficiently allocates memory space in memory-constrained systems where soft RT applications and non-RT applications compete for memory. The first objective is to balance the performance trade-off between satisfying the timing requirements of soft RT periodic tasks and optimizing the throughput of non-RT applications. The second objective is to mitigate the impacts of long stalls. \figurename~\ref{fig:approach_overview} presents the overall system architecture of SARA. First, the \textit{stall-aware dynamic memory allocation} determines appropriate memory space allocation~\encircle{2} by monitoring job execution status using our proposed PSI-based metric (i.e., $s_{intv}$)~\encircle{1}, where each job runs per period and a soft RT task consists of multiple periods. The design leverages our insight that having each soft RT job complete just before its deadline in each period fosters balancing the performance trade-off, and our observation of how $s_{intv}$, driven by memory scarcity, affects execution time. Together, these insights enable dynamic adjustment of ideal threshold and memory allocation.
    Second, a \textit{stall-aware early job dropping} mechanism is designed to mitigate the impact of unpredictable long stalls. The key idea is based on our observation of the strong correlation between the elevated $s_{intv}$ and long stall, helping us to derive an effective detection methodology~\encirclepurple{3} to proactively drop jobs at the early stage of long stall impacts~\encirclepurple{4}, thereby reducing stall time and preventing cascading performance degradation.

    \subsection{Opportunities in Performance Optimization} \label{sec:app:first}

    \textbf{Key factor in balancing performance trade-off.}
    As discussed in Section~\ref{sec:motv:motivation}, in memory-constrained systems, the performance of soft RT tasks and non-RT applications exhibits a trade-off, governed by their memory allocation, where performance enhances with the allocated memory. Therefore, to balance the trade-off, we suggest allocating \textit{just enough memory space} to soft RT tasks to ensure each job completes just before its deadline in each period, while \textit{optimizing the remaining memory} for non-RT applications. To achieve this, it is crucial to understand \textit{how memory space allocation affects execution time} during job execution. Given PSI's ability to capture the latency caused by memory shortage, we leverage PSI to understand how memory usage influences execution time.
    
    \textbf{Leveraging PSI to capture memory impact on execution time.}
    We observe that the stall time, caused by memory shortage and measured by our introduced PSI-based metric, correlates with the execution time of the soft RT job in each period. Specifically, under memory shortage, the execution time ($t_{exec}$) of a soft RT job is extended by the total stall time within a period ($s_{period}$), on top of its Best-Case Execution Time (BCET, denoted as $t_{BCET}$): 
    \begin{equation}
        t_{exec} \approx s_{period} + t_{BCET}
        \label{eq:execution_vs_PeriodStall}
    \end{equation}
    where $t_{BCET}$ represents the execution time under optimal conditions with unlimited memory resources, and $s_{period}$ is obtained by accumulating stall measurements over multiple time intervals (i.e., from $i=0$ to the total number of intervals $m$ in a period):
    \begin{equation}
        s_{period} = \sum_{i=0}^m s_{intv}
    \end{equation}
    where $s_{intv}$ is our proposed PSI-based metric, representing the maximum of memory stalls within each interval ($s_{mem}$) and I/O stalls within each interval ($s_{IO}$), both measured by PSI as absolute stall time:
    \begin{equation}
        s_{intv} = \max(s_{mem}, s_{IO})
    \end{equation}
    This formulation was chosen because memory stalls do not always fully encompass I/O stalls, contrary to the example shown in \figurename~\ref{fig:memory_and_io_stall}. This occurs because performance degradation from page faults can manifest solely as I/O stall time rather than memory stall time in some cases, as noted by TMO \cite{weiner2022tmo}. 

    \begin{figure}[hbt]
        \centering
        \includegraphics[width=0.28\textwidth]{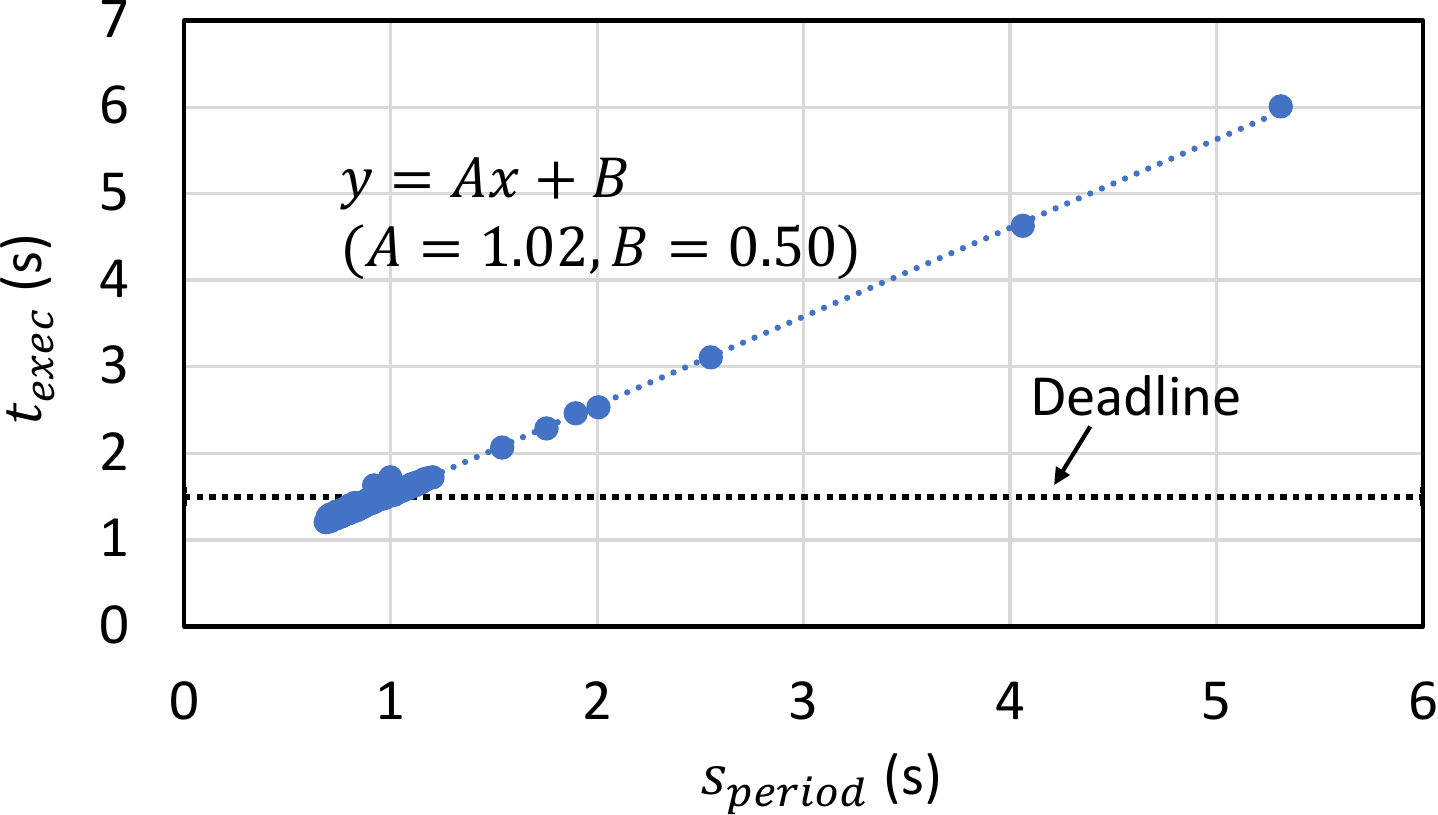}
        \setcounter{figure}{6}
        \caption{Profiling of execution time ($t_{exec}$) vs total $s_{intv}$ per period ($s_{period}$) over 500 periods.}
        \label{fig:execution_time_vs_PeriodStall_scatter_plot}
    \end{figure}

    \figurename~\ref{fig:execution_time_vs_PeriodStall_scatter_plot} provides the experimental basis for deriving Equation~\ref{eq:execution_vs_PeriodStall}. We execute a soft RT periodic task (Sphinx) for 500 periods under a constrained, fixed memory allocation while co-running a non-RT application (Graphchi). Each blue point represents the execution characteristics of the soft RT job in each period, including its execution time and $s_{period}$. Curve fitting reveals that these points exhibit a linear relationship, $y=Ax+B$, where the slope $A$ approaches $1$ and the intercept $B$ approximates 0.48, matching the BCET obtained from our profiling. This strongly supports Equation~\ref{eq:execution_vs_PeriodStall}. SARA uses this observation to derive the ideal stall time for each job to meet its deadline with just enough memory.

    \subsection{Stall-Aware Dynamic Memory Allocation} \label{sec:app:second}

\begin{figure}[hbt]
        \centering
        \includegraphics[width=0.48\textwidth]{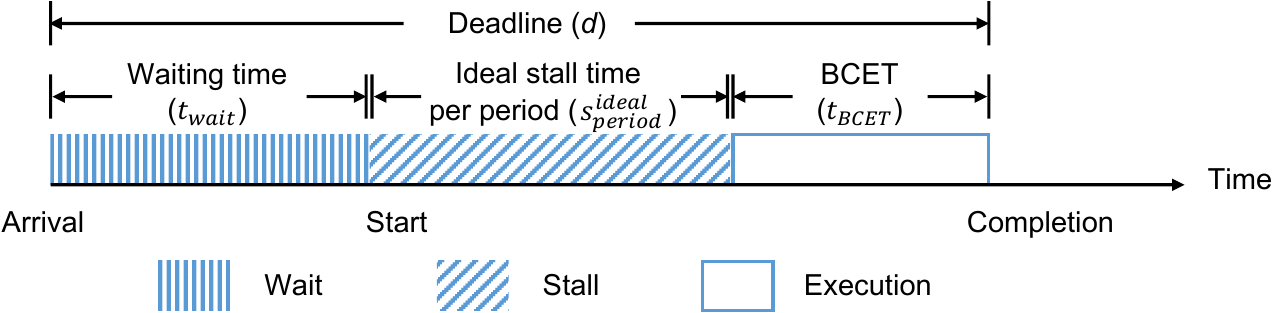}
        \caption{Relationship between deadline ($d$) and ideal stall time in a period ($s^{ideal}_{period}$).}
        \label{fig:deadline_vs_PeriodStall}
    \end{figure}

    \textbf{Deriving the ideal stall time.}
    From Equation~\ref{eq:execution_vs_PeriodStall}, we derive the ideal stall time in a period ($s^{ideal}_{period}$) for a soft RT job to complete near its deadline ($d$)
    with only the necessary amount of memory:
    \begin{equation}
        s^{ideal}_{period} = d - t_{BCET} - t_{wait}
        \label{eq:IdealPeriodStall}
    \end{equation}
    where $t_{wait}$ accounts for delays in starting the current period due to overruns from the previous period. \figurename~\ref{fig:deadline_vs_PeriodStall} illustrates this relationship. $t_{wait}$ always appears upon job arrival and can be zero or positive. Notably, stall time may be interleaved with execution time.

    To keep $s_{period}$ close to $s^{ideal}_{period}$, we must dynamically adjust memory allocation before each job completes. This is because static allocation can lead to varying $s_{period}$, as shown in \figurename~\ref{fig:execution_time_vs_PeriodStall_scatter_plot}. Additionally, adjusting memory after job completion would be ineffective, as it would either miss its deadline or already consume excessive memory, resulting in early completion. Thus, SARA monitors the stall time at a finer granularity by dynamically distributing the remaining ideal stall time across the remaining intervals, determining the ideal stall time for each monitoring interval ($s^{ideal}_{intv}$) as follows:
    
    \begin{equation}
        s^{ideal}_{intv} = \frac{s^{ideal}_{period} - \sum_{i} s^{previous}_{intv}}{{n^{remain}_{intv}}}
        \label{eq:IdealINTVStall}
    \end{equation}
    where $\sum_{i} s^{previous}_{intv}$ denotes the accumulated stall time from prior intervals. It captures the irregular runtime delays of a dynamic soft RT job and is used to update the remaining ideal stall time accordingly, thereby enabling the proposed method to adapt to dynamic workload behavior. $n^{remain}_{intv}$ represents the number of remaining monitoring intervals within a period. In this way, we use $s^{ideal}_{intv}$ to guide memory allocation at each monitoring interval.

    \textbf{Dynamic memory allocation.}
    To ensure each soft RT job completes close to its deadline, we adjust memory allocation at each interval to make $s_{intv}$ approach $s^{ideal}_{intv}$, defined in Equation~\ref{eq:IdealINTVStall}. Therefore, when $s_{intv}$ exceeds $s^{ideal}_{intv}$, indicating a potential deadline miss, more memory is allocated to the current job; otherwise, memory is reclaimed to improve the throughput of non-RT applications. The memory adjustment amount is calculated as follows:
    \begin{equation}
    mem_{alloc} = x \times \frac{s_{intv} - s^{ideal}_{intv}}{l_{intv}}
    \end{equation}    
    where $x$ defines the base adjustment unit of memory allocation. Larger deviations between $s_{intv}$ and $ s^{ideal}_{intv}$ result in proportionally larger memory adjustments, which are normalized by the monitoring interval length ($l_{intv}$) for consistent scaling.
    
    By identifying opportunities for performance optimization and deriving an appropriate threshold ($s^{ideal}_{intv}$), which is dynamically updated based on runtime conditions, SARA effectively enables soft RT tasks to meet deadlines with just enough memory, thereby optimizing the throughput of non-RT applications.

    \subsection{Stall-Aware Early Job Dropping} \label{sec:app:third}

    \begin{figure}[hbt]
         \centering
         \begin{subfigure}[b]{0.235\textwidth}
             \centering
             \includegraphics[width=\textwidth]{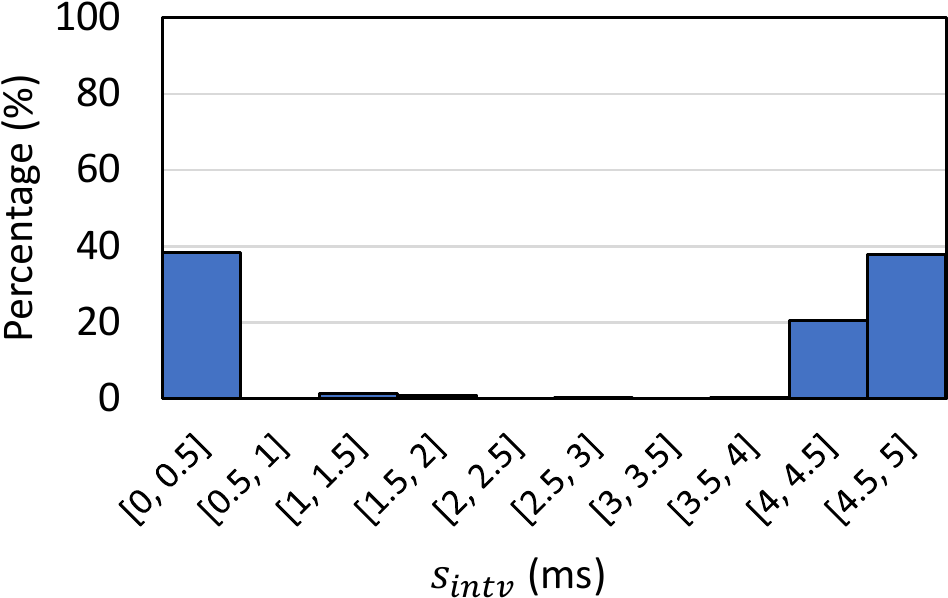}
             \caption{Normal period.}
             \label{fig:image1}
         \end{subfigure}
         \begin{subfigure}[b]{0.235\textwidth}
             \centering
             \includegraphics[width=\textwidth]{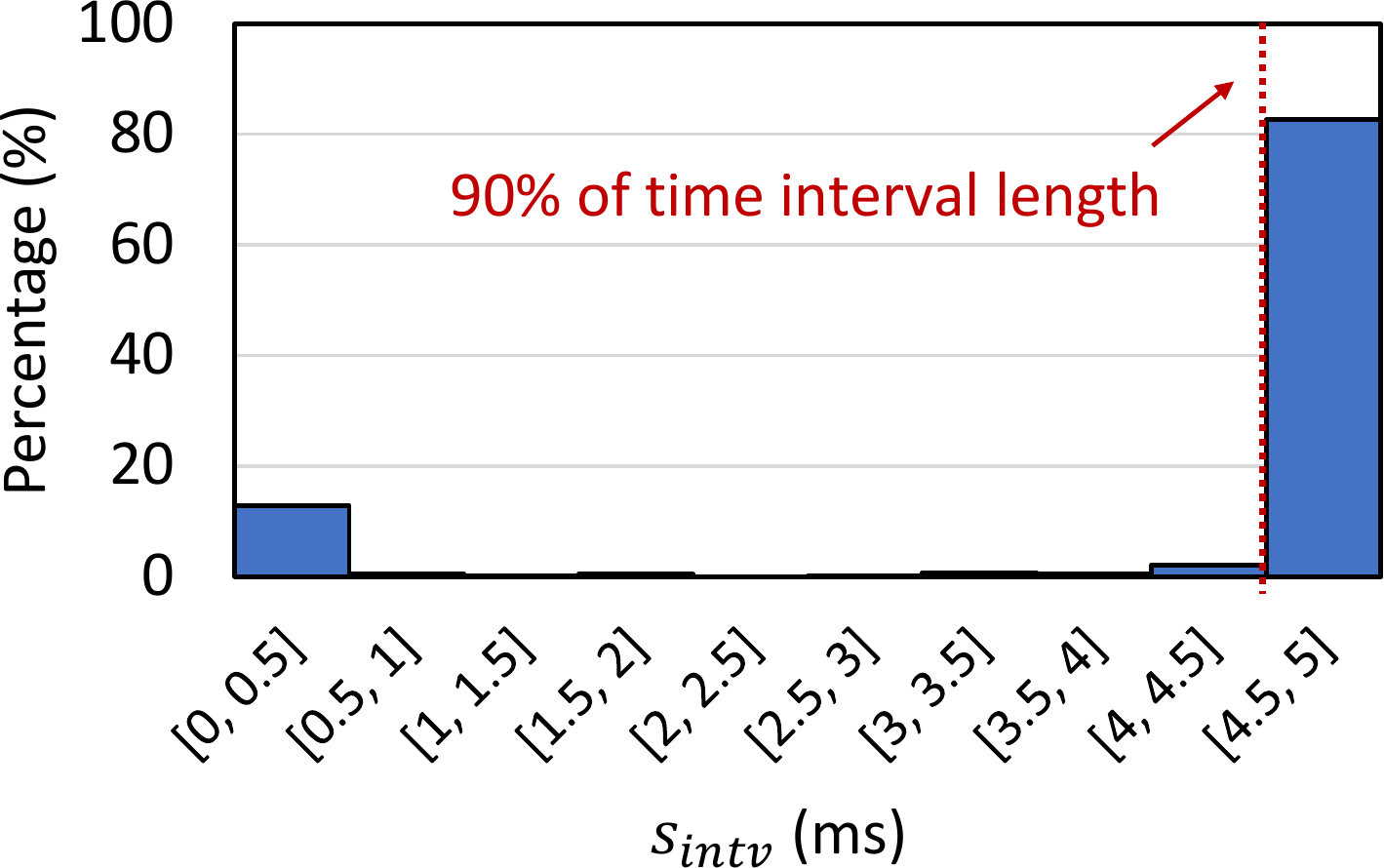}
             \caption{Long stall occurred period}
             \label{fig:image2}
         \end{subfigure}
        \caption{$s_{intv}$ distribution of normal and long stall periods.}
        \label{fig:INTVStall distribution}
    \end{figure}
    \begin{figure}[hbt]
        \centering
        \begin{subfigure}[b]{0.32\textwidth}
            \centering
            \includegraphics[width=\textwidth]{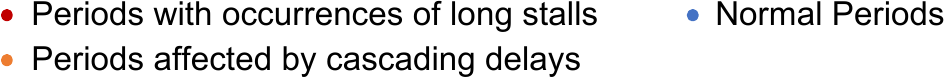}
            \vspace{-2mm}
            \label{fig:long_stall_analysis_label}
        \end{subfigure}
        \hfill
        \begin{subfigure}[b]{0.24\textwidth}
            \centering
            \includegraphics[width=\textwidth]{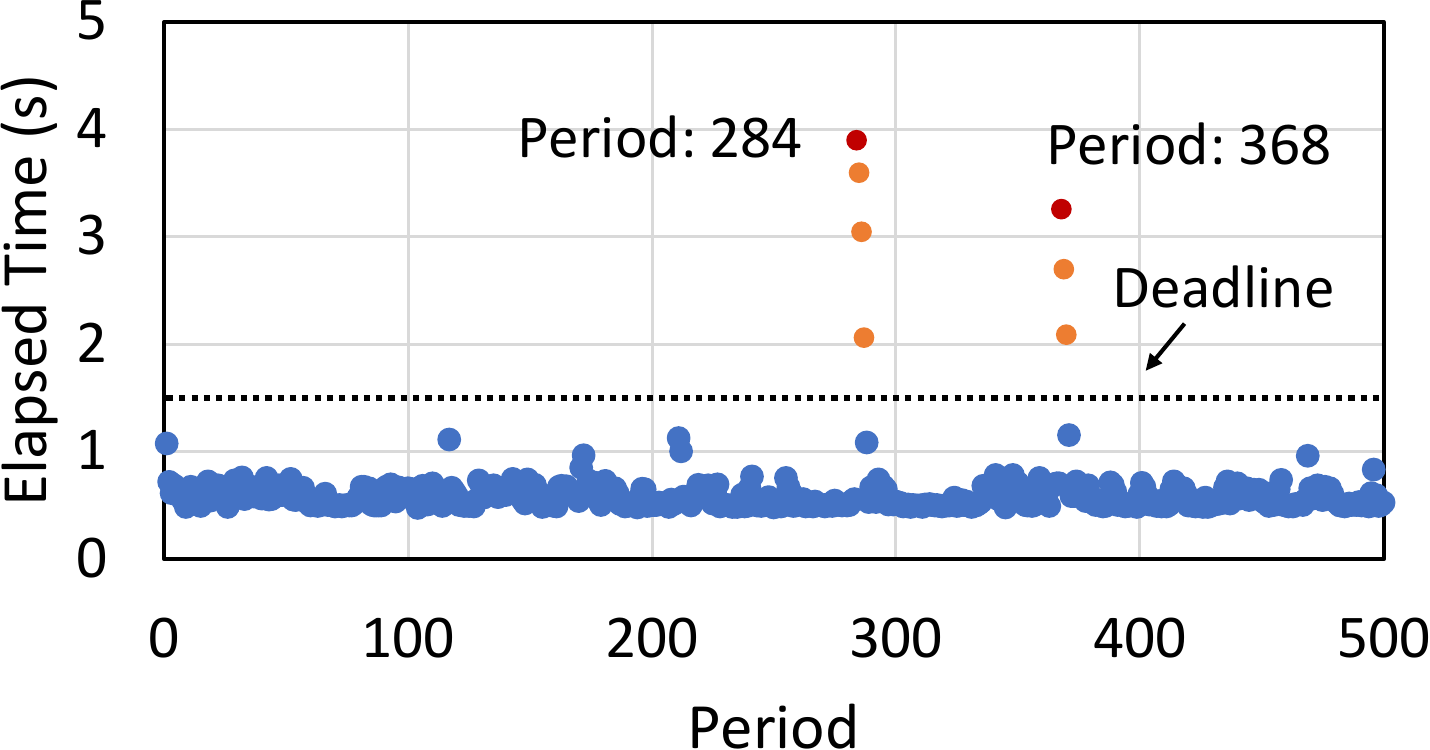}
            \caption{Elapsed time of normal and long stall-affected periods.}
            \label{fig:long_stall_analysis_1}
        \end{subfigure}
        \hfill
        \begin{subfigure}[b]{0.24\textwidth}
            \centering
            \includegraphics[width=\textwidth]{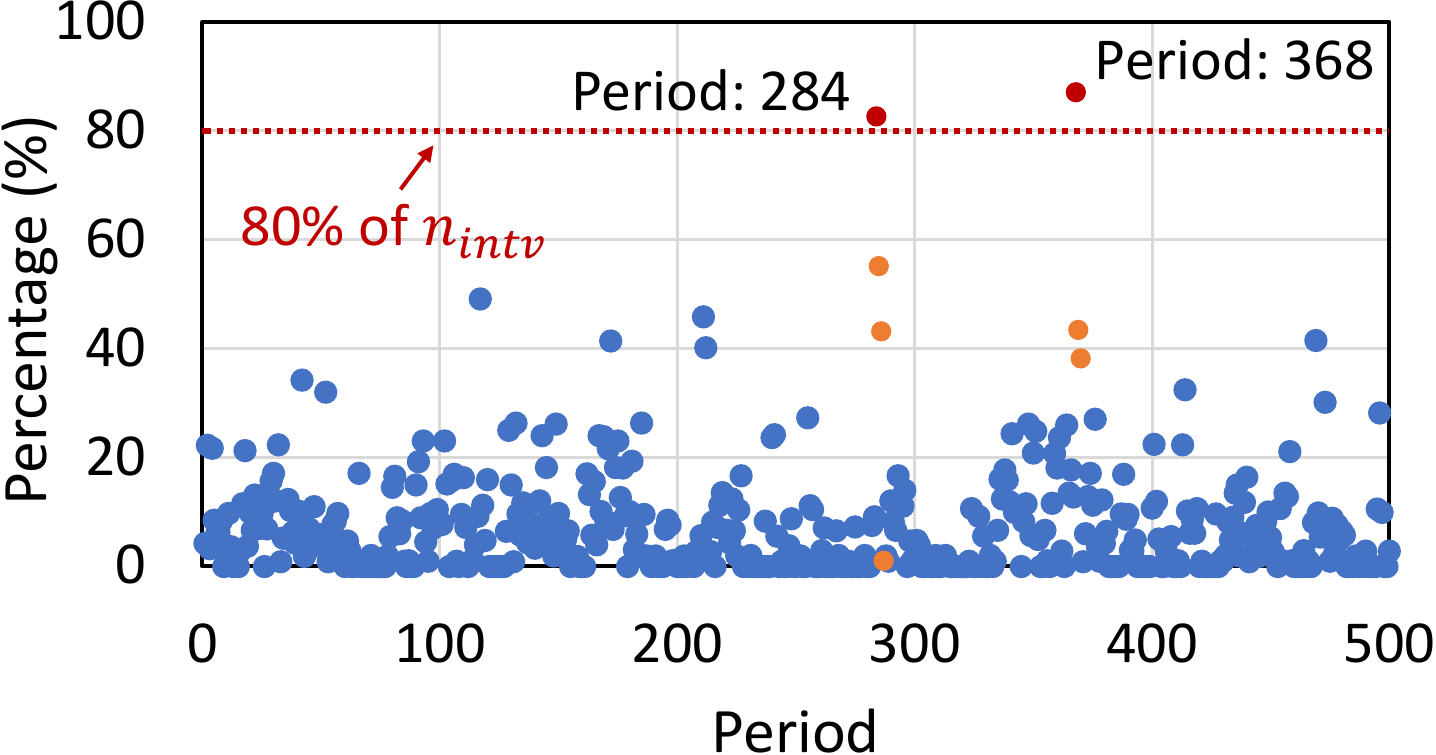}
            \caption{Percentage of $n_{intv}$ with $s_{intv}$ exceeding 90\% of the $l_{intv}$.}
            \label{fig:long_stall_analysis_2}
        \end{subfigure}
        \caption{The correlation between elapsed time and analysis results.}
        \label{fig:INTVStall_distribution}
    \end{figure}

    \begin{figure*}[t]
        \centering
        \begin{subfigure}[b]{0.491\textwidth}
            \centering
            \includegraphics[width=\textwidth]{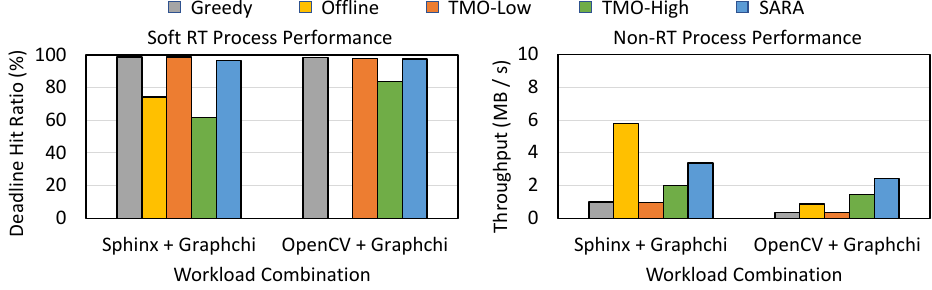}
            \caption{Workload performance in the small-memory, slow-SSD system.}
            \label{fig:evaluation_slow_small}
        \end{subfigure}
        \hspace{0.005\textwidth}
        \begin{subfigure}[b]{0.491\textwidth}
            \centering
            \includegraphics[width=\textwidth]{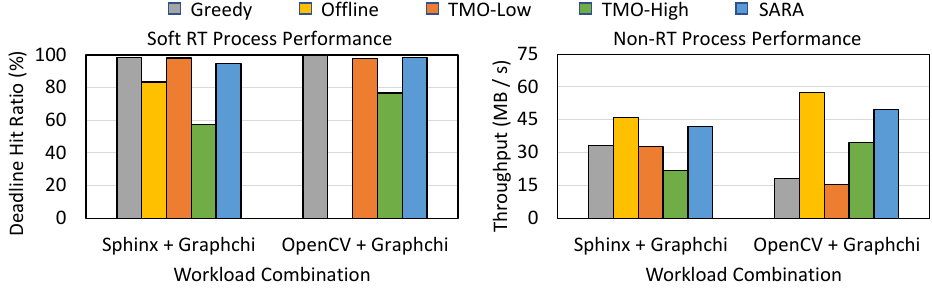}
            \caption{Workload performance in the large-memory, slow-SSD system.}
            \label{fig:evaluation_slow_large}
        \end{subfigure}

        \vspace{0.02\textwidth}

        \begin{subfigure}[b]{0.491\textwidth}
            \centering
            \includegraphics[width=\textwidth]{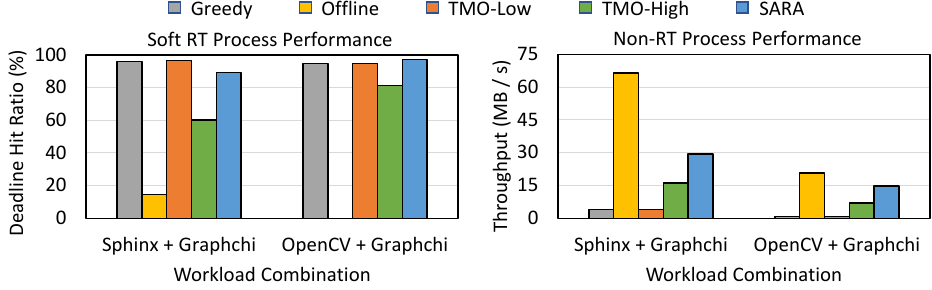}
            \caption{Workload performance in the small-memory, fast-SSD system.}
            \label{fig:evaluation_fast_small}
        \end{subfigure}
        \hspace{0.005\textwidth}
        \begin{subfigure}[b]{0.491\textwidth}
            \centering
            \includegraphics[width=\textwidth]{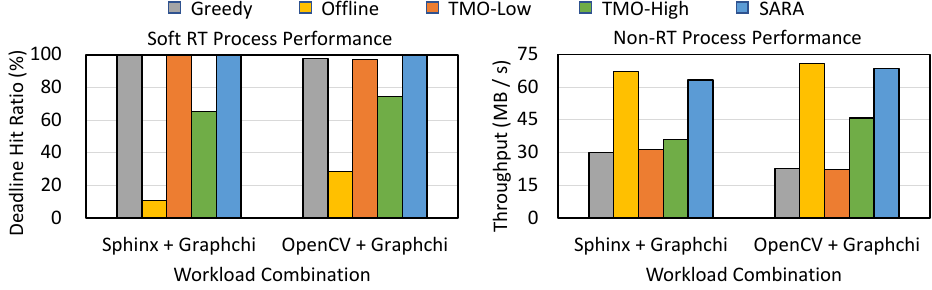}
            \caption{Workload performance in the large-memory, fast-SSD system.}
            \label{fig:evaluation_fast_large}
        \end{subfigure}
        
        \caption{Performance of soft RT and non-RT workloads under different management approaches on diverse system configurations.}
        \label{fig:performance}
    \end{figure*}
    
    \textbf{Defining long stalls.}
    Timely mitigation of the impact of unpredictable long stalls requires a well-defined criterion to identify their occurrence. \figurename~\ref{fig:INTVStall distribution} shows distinct $s_{intv}$ distribution characteristics between normal and long stall occurred periods of a soft RT periodic task (Sphinx). In the long stall period, the distribution shows that most $s_{intv}$ values cluster at high magnitudes (i.e., 4.5–5 ms, where 5 ms is the interval length $l_{intv}$), in contrast to the normal period. This pattern reveals the trend of long stalls, which typically occur when a high percentage of intervals encounter severe stalls. \figurename~\ref{fig:INTVStall_distribution} shows how SARA determines the percentage of intervals with severe stalls that qualify as long stalls.
    Long stall periods (marked in red) exhibit over 80\% of intervals with severe stalls (i.e., $s_{intv}$ over 90\% of $l_{intv}$), which distinguishes them from normal periods. 
    Therefore, we define long stalls as follows.

    \begin{definition} \label{def:longstall}
        The long stall is defined as when at least $m\%$ of the total monitoring intervals experience a stall time $s_{intv}$ exceeding $n\%$ of the interval length $l_{intv}$.
    \end{definition} 

    We empirically select the values of $m$ and $n$ from a range of candidates, as long as they effectively distinguish long stall periods from normal ones. These values are determined via profiling only once for each soft RT workload on each system, thereby avoiding additional overhead in subsequent memory management. Moreover, soft RT periodic tasks typically follow similar control paths, allowing the profiling results to reliably detect long stalls.


    \textbf{Detecting long stalls and early job dropping.}
    To prevent cascading delays caused by long stalls, we proactively drop running jobs early if they encounter long stalls, as determined by Definition~\ref{def:longstall}.
    To drop jobs as early as possible while ensuring that only those unable to meet deadlines are dropped, we detect long stalls when the remaining ideal stall time within the execution period falls below 0 (i.e., $s^{ideal}_{period} - \sum_{i} s^{previous}_{intv} < 0$). This condition can occur before deadlines, allowing early job dropping. Additionally, it indicates inevitable deadline misses due to the lack of additional tolerable stall time. When long stalls are detected, jobs are dropped to prevent cascading delays for jobs in subsequent periods, while also freeing memory for non-RT applications to enhance throughput.
    
\section{Performance Evaluation} \label{sec:eval}
\subsection{Experimental Setup} \label{sec:setup}

    \textbf{System configuration.} We conducted our experiments on a real system with an Intel\textsuperscript{®} Core\textsuperscript{TM} i7-13700 (16 cores, 2.10 GHz) CPU with 64 GB DDR5 RAM. For the swap backend, we test on two SSDs---\textit{one slow and one fast}---to demonstrate SARA’s flexibility across heterogeneous storage backends. Specifically, we used a standard SSD (560 MB/s read, 510 MB/s write) and a high-performance SSD (2100 MB/s read, 1000 MB/s write). The system ran Linux kernel 6.5.0 with cgroup v2. Note that to simulate memory-constrained conditions, we use cgroups to limit memory size to 60\% (\textit{small-memory}) and 80\% (\textit{large-memory}) of the aggregate peak memory usage of workloads.

\renewcommand{\arraystretch}{1.2}    
\begin{table}[hbt]
    \centering
    \setlength{\abovecaptionskip}{2pt}
    \setlength{\belowcaptionskip}{0pt}
    \caption{Workload characteristics.}
    \label{tab:workload_char}
    \begin{tabular}{|c|c|c|c|c|}
        \hline
        \multicolumn{2}{|c|}{Workload} & Sphinx & OpenCV & Graphchi \\ \hline
        \multicolumn{2}{|c|}{Type} & Soft RT & Soft RT & Non-RT \\ \hline
        \multicolumn{2}{|c|}{Deadline (s)} & 1.5 & 0.8 & N/A \\ \hline
        \multicolumn{2}{|c|}{BCET (s)} & 0.48 & 0.14 & N/A \\ \hline
        \multirow{3}{*}{\shortstack{Memory\\Usage\\(MB)}} & Max & 338 & 183 & 272 \\ \cline{2-5}
         & Avg & 281 & 166 & 244 \\ \cline{2-5}
         & Min & 136 & 97  & 36  \\ \hline
    \end{tabular}
\end{table}

    \textbf{Workload.} We select two representative soft RT workloads from edge AI applications, considering different memory usage patterns: Sphinx~\cite{thomas2014cortexsuite}, a speech recognition application with high memory usage, and OpenCV~\cite{itseez2014theopencv}, an image classification application with low memory usage. These soft RT workloads are executed with Graphchi~\cite{kyrola2012graphchi}, a non-RT, memory-intensive workload that processes and manages graphs. The characteristics of the workloads are described in \tablename~\ref{tab:workload_char}.
    
    \textbf{Parameters.} The monitoring interval length, $l_{intv}$, is set on the millisecond scale (e.g., 5~ms), as modern systems typically use a kernel tick rate between 100~Hz and 1000~Hz~\cite{TickRate}. These kernel timer ticks trigger task switching and may cause stall state transitions, which in turn affect PSI values. Thus, we adopted the millisecond scale to effectively capture these dynamics.
    The base adjustment unit of memory allocation, $x$, is set to a small number of megabytes (e.g., 1~MB), considering that the memory usage of workloads is on the order of hundreds of megabytes. This corresponds to less than 1\% of the workloads' memory usage, which helps avoid abrupt performance degradation and memory overprovisioning.
    Based on the profiling results, $m$ and $n$ are set to 80 and 90 for both Sphinx and OpenCV.\footnote{These parameters are applied for systems with slow SSD, as no long stalls occur on fast SSDs.}

    \textbf{Baseline.} We evaluated SARA against four memory allocation approaches: Greedy~\cite{memory_sufficiency}, Offline~\cite{delimitrou2014quasar,sfakianakis2018quman,zhu2022qos}, TMO-Low and TMO-High~\cite{weiner2022tmo}. Greedy allocates memory based on the actual usage of soft RT workloads without constraints. Offline determines static memory allocation through offline profiling to meet real-time requirements. TMO-Low, the original TMO parameter setting, uses a conservative 0.1\% PSI threshold, while TMO-High uses a higher threshold determined through exhaustive search\footnote{As TMO is unaware of the relationship between real-time requirements and PSI, it is unsuitable for identifying adaptive thresholds. Thus, we resort to exhaustive search to determine a static one.}, bringing the average elapsed time of soft RT tasks just below deadlines. In all these approaches, the non-RT workload uses the remaining memory after allocation for soft RT workloads.



        

\subsection{Experimental Result} \label{sec:result}
\textbf{Overall discussion.}
    \figurename~\ref{fig:performance} shows the deadline hit ratios of soft RT tasks and the throughput of the non-RT application on our experimental platform under small and large memory configurations with slow and fast SSD backends. SARA achieves a near-optimal average deadline hit ratio of about 97.13\%. Meanwhile, compared to Greedy and TMO-Low, which also maintain high deadline hit ratios, SARA improves throughput by 5.69$\times$ and 5.89$\times$, respectively. Notably, the improvement is particularly pronounced in the small-memory, fast-SSD system, reaching up to 20.96$\times$ and 22.32$\times$ over Greedy and TMO-Low, respectively.
    The performance benefits mainly stem from dynamically allocating just enough memory to meet real-time requirements while optimizing the remaining memory for the non-RT application, as well as early job dropping to mitigate long stalls. In the small-memory, fast-SSD system, the non-RT application especially benefits from additional memory, as it processes large graphs that require sufficient memory. Moreover, the high I/O bandwidth of the fast SSD allows soft RT tasks to use less memory to meet deadlines, while the non-RT application takes advantage of both the additional memory and increased I/O bandwidth.
    
    Although Offline and TMO-High achieve relatively high throughput, they have deadline hit ratios of only 49.61\% and 77.11\%, respectively. This severe failure to meet the timing requirements of soft-RT tasks is unacceptable. Offline allocates static memory without considering the runtime dynamicity of systems, leading to significant variance in elapsed time. Jobs with longer elapsed times can delay subsequent jobs, potentially resulting in a 0\% deadline hit ratio. TMO-High uses a static high threshold that is not adjusted for system variance, causing jobs to miss deadlines or complete early, preventing full optimization of throughput. On the other hand, both Greedy and TMO-Low successfully maintain real-time guarantees; however, they result in significantly low throughput. Greedy allocates unrestricted memory to soft RT tasks, which restricts memory available to the non-RT application, leading to excessive page swapping and degraded throughput. Similarly, TMO-Low allocates excessive memory to soft RT tasks to keep PSI near the low threshold, at the cost of greatly reduced non-RT application throughput. These results highlight that, even with limited memory, SARA effectively satisfies real-time requirements while simultaneously optimizing non-RT application throughput through efficient memory allocation.

\begin{figure}[hbt]
         \centering
         \begin{subfigure}[b]{0.24\textwidth}
             \centering
             \includegraphics[width=\textwidth]{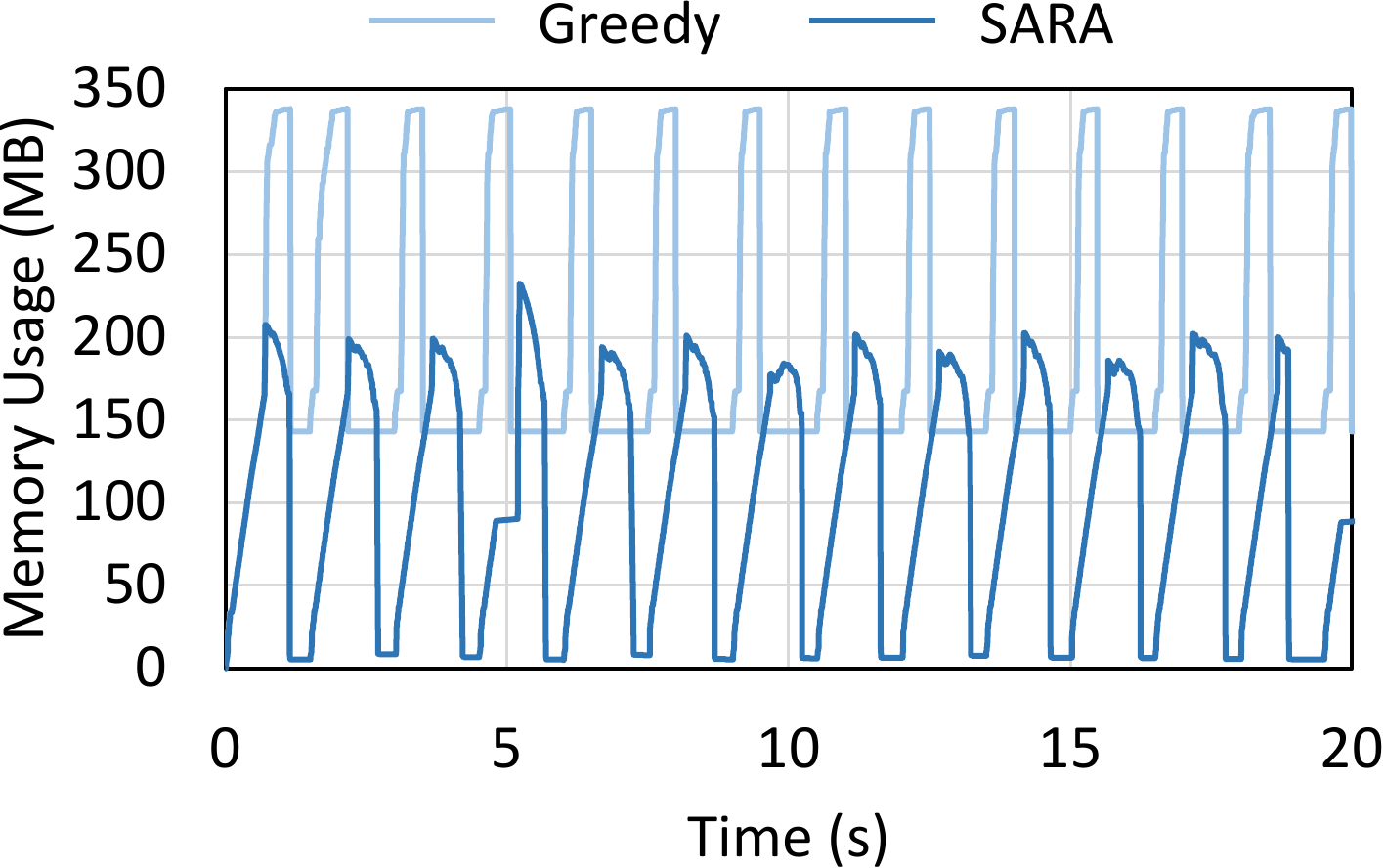}
             \caption{Memory usage of Sphinx.}
             \label{fig:image1}
         \end{subfigure}
         \begin{subfigure}[b]{0.24\textwidth}
             \centering
             \includegraphics[width=\textwidth]{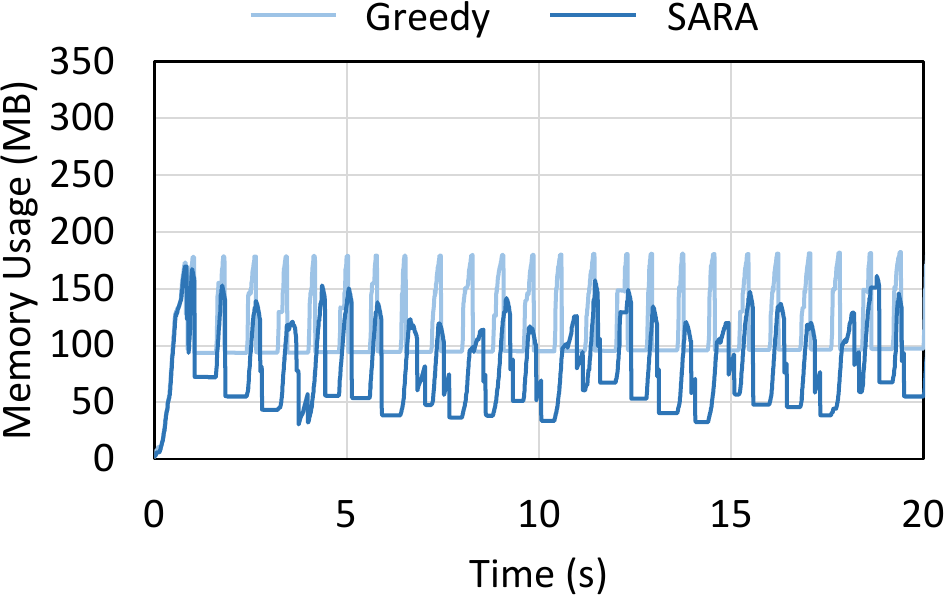}
             \caption{Memory usage of OpenCV.}
             \label{fig:image2}
         \end{subfigure}
        \caption{Memory usage reduction of soft RT periodic tasks with stall-aware dynamic memory allocation.}
        \label{fig:reduction_of_memory_savings}
    \end{figure}

    \textbf{Effectiveness of stall-aware dynamic memory allocation.} \figurename~\ref{fig:reduction_of_memory_savings} compares memory usage patterns of Sphinx and OpenCV when running with Graphchi under SARA and Greedy approaches. It is clear that the stall-aware dynamic memory allocation component enables the soft RT job in each period to use less memory to meet its deadline by dynamically reducing memory allocation when it can tolerate longer stall time. This frees up memory for non-RT applications, significantly improving throughput.

    \begin{figure}[hbt]
         \centering
         \begin{subfigure}[t]{0.24\textwidth}
             \centering
             \includegraphics[width=\textwidth]{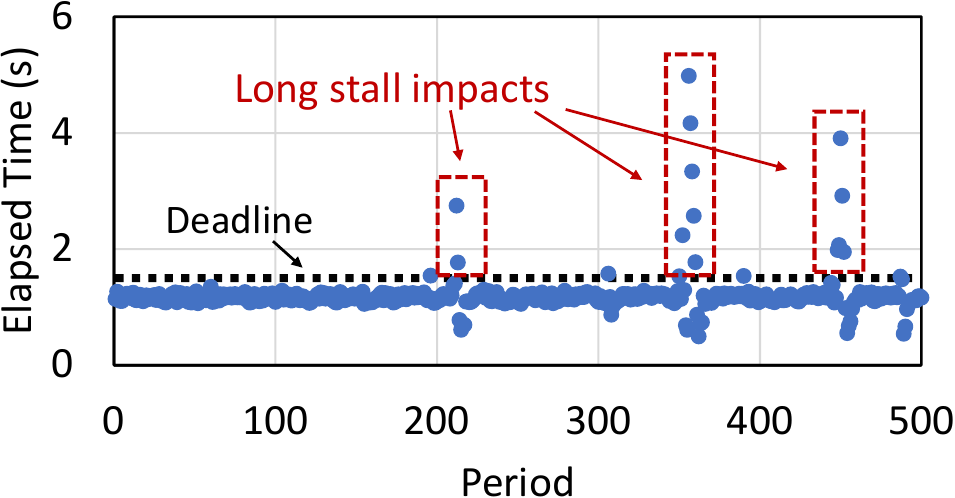}
             \caption{Elapsed time of each job over 500 periods under SARA, without stall-aware early job dropping.}
             \label{fig:image1}
         \end{subfigure}
         \hfill
         \begin{subfigure}[t]{0.24\textwidth}
             \centering
             \includegraphics[width=\textwidth]{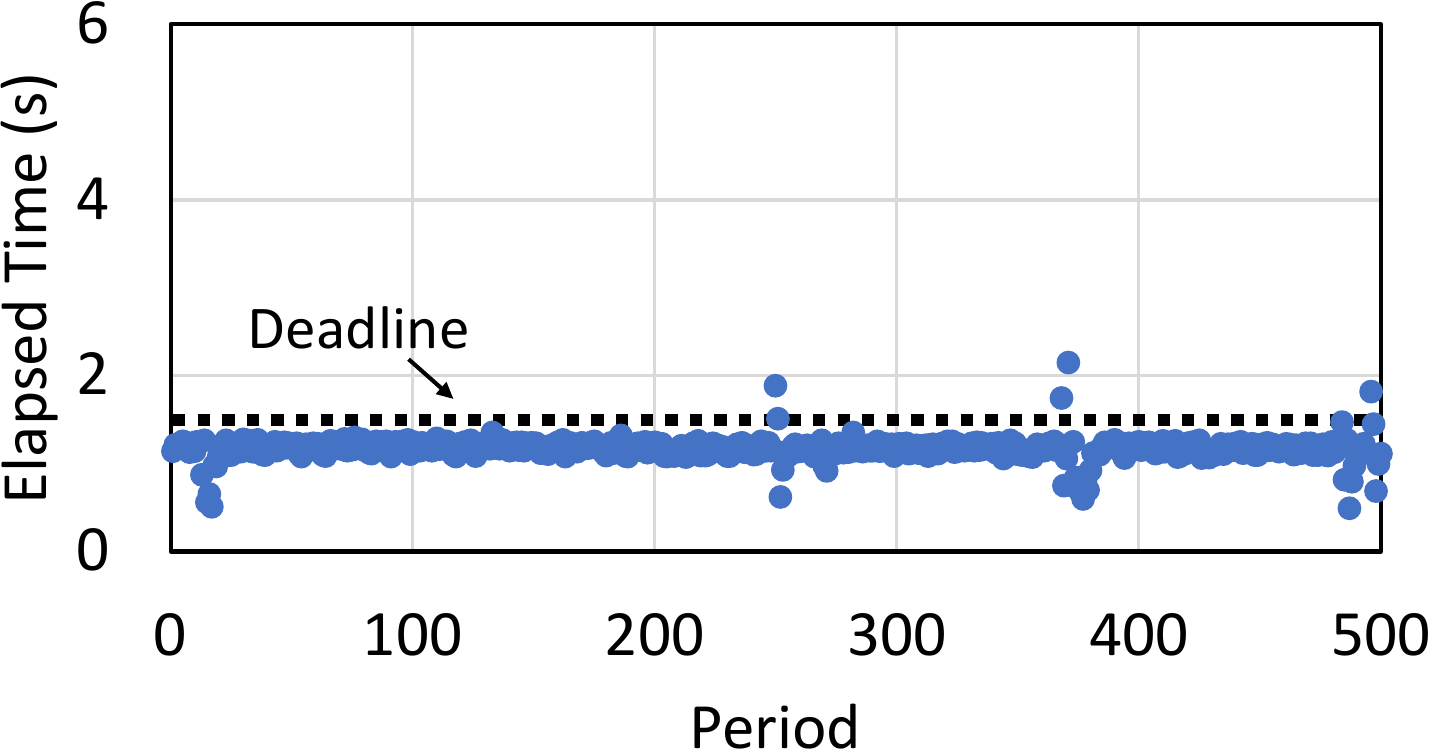}
             \caption{Elapsed time of each job over 500 periods under SARA.\newline}
             \label{fig:image2}
         \end{subfigure}
        \caption{Reduction of long stall impacts on Sphinx with stall-aware early job dropping.}
        \label{fig:reduction_of_long_stall_impacts}
    \end{figure}

    \textbf{Effectiveness of stall-aware early job dropping.}
    \figurename~\ref{fig:reduction_of_long_stall_impacts} presents the elapsed time of Sphinx jobs over 500 periods under SARA with and without the stall-aware early job dropping. The figure shows that SARA reduces the occurrence of prolonged jobs and mitigates cascading delays by detecting and dropping stalled jobs in the early stage. SARA improves the deadline hit ratio by 1.2\% compared to the scenario without early job dropping, although dropping operations take time and may cause some jobs to slightly exceed their deadlines.


\section{Related Work} \label{sec:rel}
    Memory management strategies typically fall into two categories: page pinning~\cite{lee2012mrt,wu2023app}, and memory space allocation through cgroups~\cite{memory_sufficiency, delimitrou2014quasar,lagar2019software, sfakianakis2018quman, weiner2022tmo, zhu2022qos}. Page pinning retains frequently accessed pages in memory for soft RT tasks by modifying the Linux page replacement policy. However, this fine-grained management incurs high management overhead and lacks adaptability to OS updates.
    In contrast, memory space allocation approaches focus on allocating memory space without managing individual pages, offering better cost-effectiveness and adaptability to system updates. Among memory allocation approaches, current enterprise solutions \cite{memory_sufficiency} greedily allocate memory to soft RT tasks to meet timing requirements. The offline profiling approaches \cite{delimitrou2014quasar,sfakianakis2018quman,zhu2022qos} estimate the required memory for achieving the expected performance for processes based on offline results. Lagar-Cavilla et al. \cite{lagar2019software} introduce page swapping-in rate as an online metric for allocation decisions. However, we find that these works rely on unsuitable metrics, failing to achieve efficient allocation in mixed-criticality systems. Recent work TMO~\cite{weiner2022tmo} introduces a new metric, PSI, that directly measures processes experienced stall time due to resource shortage across memory and I/O via the cgroup interface. Nevertheless, TMO remains inadequate for efficient memory allocation in mixed-criticality systems, as it lacks a mechanism to identify adaptive PSI thresholds for accurately monitoring soft RT execution status. Our proposed SARA is the first to establish the relationship between memory shortages and execution time through our proposed PSI-based metric. This relationship enables SARA to achieve our proposal of allocating just enough memory for soft RT tasks to meet deadlines and mitigate long stalls in memory-constrained systems, thus meeting real-time requirements and improving throughput.

\section{Conclusion} \label{sec:conc}
    We introduce SARA, a novel stall-aware real-time memory allocator designed to balance the trade-off between real-time requirements and throughput in memory-constrained systems. SARA dynamically and efficiently allocates just enough memory to soft RT tasks to meet deadlines while optimizing the remaining memory for non-RT applications. This allocation is guided by the ideal stall time, derived from our insight into how memory impacts execution time. It also mitigates long stalls by proactively dropping prolonged jobs, detected based on our definition. SARA maintains near-optimal deadline hit ratios for soft RT tasks while improving non-RT application throughput by up to 22.32$\times$ compared to existing approaches, even when memory capacity is limited to 60\% of peak demand.

\section*{Acknowledgement}
    This work was supported in part by National Science and Technology Council under grant nos. 111-2221-E-002-152-MY3, 112-2221-E-002-160-MY3, 114-2927-I-002-525, 114-2927-I-002-532, 114-2223-E-002-011, 114-2221-E-002-219-MY3, 114-2221-E-002-222-MY3, and by Ministry of Education under grant no. 114L900903.

\bibliographystyle{IEEEtran}
\bibliography{reference}

\end{document}